\documentclass[11pt,a4paper]{article}

\usepackage[T1]{fontenc}
\usepackage[latin2]{inputenc}
\usepackage{graphicx}
\usepackage[normalem]{ulem}

\usepackage{fancyhdr}
\usepackage{calc}
\usepackage{epsfig}
\usepackage{amssymb}
\usepackage{amsmath}
\usepackage{indentfirst}
\usepackage{chngcntr}
\usepackage{jheppub}

\def\lc{l}
\def\sp{S}
\def\s0{\sigma_0}

\title{Twist expansion of Drell-Yan structure functions in color dipole approach}
\author[a]{Leszek Motyka}
\author[a]{Mariusz Sadzikowski}
\author[a]{Tomasz Stebel}
\affiliation[a]{Institute of Physics, Jagiellonian University\\
S. \L{}ojasiewicza 11,  30-348 Krak\'{o}w, Poland}
\emailAdd{leszek.motyka@uj.edu.pl}
\emailAdd{mariusz.sadzikowski@uj.edu.pl}
\emailAdd{tomasz.stebel@uj.edu.pl}
\abstract{
The forward Drell-Yan process at the LHC probes the proton structure at a very small Bjorken-$x$ and moderate hard scales. In this kinematical domain higher twist effects may be significant and introduce sizeable corrections to the standard leading twist description. We study the forward Drell-Yan process beyond the leading twist approximation within the color dipole model framework that incorporates multiple scattering effects. We derive the Mellin representation of the forward Drell-Yan impact factors for fully differential cross-sections. These results combined with the color dipole cross-section of the saturation model are used to perform the twist expansion of the Drell-Yan structure functions at arbitrary transverse momentum $q_T$ of the Drell-Yan pair and also of the structure functions integrated over $q_T$. We also investigate the Lam-Tung relation, find that it is broken at twist 4 and provide explicit estimates for the breaking term.}
\arxivnumber{1412.4675}

\begin{document}
\maketitle

\section{Introduction and conclusions}

The forward Drell-Yan (DY) processes at the LHC are expected to provide the most sensitive measurements of  parton densities in the proton down to very small $x \simeq 10^{-6}$. At the LHCb experiment the DY lepton--antilepton pair may be measured down to invariant mass $M$ of about 2.5~GeV, so the related parton density to the scale $\mu^2 \simeq 6.25$~GeV$^2$ \cite{lhcb0}. This kinematic region has been never probed before. It extends HERA measurements of proton structure at moderate scales towards small parton~$x$ by about two orders of magnitude. In fact, the forward Drell-Yan process is a unique tool to explore this region. Hence, it is mandatory to acquire deep theoretical understanding of the process in QCD.

The region of moderate scales and very small~$x$ is sensitive to interesting QCD effects. The standard DGLAP description of parton densities evolution in the proton may be significantly affected by small-$x$ resummation effects \cite{bfkl,bfklrev} and higher twist contributions related (but not identical) to multiple scattering corrections. Therefore, the forward Drell-Yan process may be used as a sensitive probe of these effects. On the other hand, a good theoretical understanding of higher twist corrections to the Drell-Yan cross-section is necessary to extract the standard twist-2 parton densities with higher precision and reduced uncertainties. Thus, in this paper we address the problem of higher twist effects in the forward Drell-Yan processes.

The forward Drell-Yan process has multiple advantages as the probe of the proton structure and QCD dynamics. The presence of hard electromagnetic probe allows for effective application of perturbative QCD. From the experimental side, the kinematic variables of the final state composed of a lepton--antilepton pair may be measured with a good precision, giving access to multiple differential distributions. In particular the Drell-Yan pair angular distributions are determined by four invariant structure functions, $T_i$, $i=1,\ldots,4$, describing proton interactions with a virtual photon which mediates the lepton pair production \cite{LamTung1,LamTung2}. All structure functions $T_i$ may be decomposed into twist-series using the Operator Product Expansion (OPE), in which the leading twist-two contributions may be computed using standard parton densities and the (unknown) higher twist terms are suppressed by negative powers of the process hard scale. This power suppression, however, may be compensated in the region of moderate scale and very small-$x$ by rapidly growing higher twist matrix elements, so that the higher twist contributions are expected to become important below $\mu^2 \sim 30$~GeV$^2$ \cite{GBLS}.\footnote{Existing results on twist-$\tau$ evolution at small-$x$ indicate that leading twist-$\tau$ gluonic matrix elements grow with decreasing-$x$ faster than powers $\geq \tau/2$ of the large gluon density $xg(x,\mu^2)$.} Hence the four DY structure functions carry enriched information on higher twists: the higher twist hadronic matrix elements are coupled to four different coefficient functions. This gives opportunities for broader and more detailed tests of the higher twist description and provides tools for more efficient isolation of higher twist corrections. In particular, it follows from the famous Lam-Tung relation \cite{LamTung2} that certain combination of the DY structure functions vanishes at twist-2 (up to next-to-next-to leading order corrections), and therefore its deviation from zero is a sensitive probe of higher twist effects.

As yet the higher twist contributions to the proton structure and to the Drell-Yan structure functions are poorly known from experiment and the rigorous theoretical description of the higher twist terms within QCD is highly involved. A treatment of the higher twist contribution in DY scattering within collinear QCD proposed in Ref.~\cite{QS}, see also \cite{FMSS,FSSM}, still requires modeling of higher twist matrix elements. A practical way to circumvent these obstacles is to adopt, as a first approximation, the eikonal or Glauber-Mueller picture where multiple scattering in QCD is a product of independent single scatterings. This approach was implemented e.g.\ in the very successful Golec-Biernat--W\"{u}sthoff (GBW) saturation model \cite{GBW}.

 The GBW model was proved to provide an efficient and accurate unified picture of multiple HERA  processes at small~$x$: deeply inelastic scattering (DIS), diffractive DIS, elastic vector meson production, deeply virtual Compton scattering \cite{GBW,KMW}. In the high energy limit in QCD, relevant for the forward Drell-Yan processes at the LHC, the DY scattering amplitudes may be computed using the high-energy factorization ($k_T$-factorization framework \cite{GLR}) and the transverse position space. This leads to the `color dipole picture' \cite{NZ} of the forward DY process, proposed by \cite{Brodsky,Kopeliovich} in which the QED-QCD partonic amplitudes of virtual photon production are combined with the color dipole cross-section, that needs to be fitted and/or modeled (beyond the leading order contribution at the leading twist). The dipole cross-section is constrained by the HERA data, and the fit of results may be applied to predict the DY cross-sections including the higher twist effects \cite{Brodsky,Kopeliovich,GJ,GBLS,Ducati}, see also Ref.\ \cite{BMS}. Thus, we shall estimate the higher twist effects in the structure functions of the forward DY processes using the color dipole approach and the GBW saturation model. One should stress that at the leading twist the dipole approach is consistent with the standard collinear picture results in the high energy limit up to the NLO, but it provides in addition an estimate of the multiple scattering and higher twist effects.

In this paper the cross-section decomposition into its twist components is carried out in the Mellin representation for the color dipole sizes. This decomposition method was initially proposed in Ref.~\cite{BGBP}, further developed in Ref.~\cite{BGBM} and then applied to the total forward DY cross-section \cite{GBLS} and diffractive DIS~\cite{MMS}. In this framework the twist contributions are related to complex singularities (poles or branch points) in the Mellin plane. We follow the technique of the total DY cross-section twist decomposition proposed in \cite{GBLS}, but extend the results to differential cross-sections in the lepton angles and the DY pair transverse momentum $q_T$. In more detail we compute the Mellin transforms of all the forward DY impact factors at given $q_T$ and combine the results with the color dipole cross-section to get $q_T$-dependent twist decomposition of all the four DY structure functions. Here we restrict ourselves to the simplest GBW eikonal model of the color dipole cross-section but it is straightforward to combine the obtained Mellin representation of the impact factors with other descriptions or parameterizations of the color dipole cross-section.

The findings of this paper may be summarized as follows. The key novel result are the Mellin representations of the forward DY impact factors for all the DY structure functions at an arbitrary transverse momentum $q_T$. These impact factors follow directly from perturbative QCD. The impact factors are then combined with the GBW color dipole cross-section to get an analytic form of the twist expansion of the DY structure functions. These results are model dependent but they exhibit some generic features driven by the perturbative part, like e.g.\ the saturation of the Lam-Tung relation at twist~2, the presence or absence of hard scale logarithms. We obtain results both for the helicity structure functions and for the invariant structure functions. We find that the Lam-Tung relation holds at twist~2 but it is broken at twist~4, and the breaking term is leading in perturbative QCD at this twist. Thus, the Lam-Tung combination of the DY structure functions is a promising observable for experimental measurements of the higher twist effects. In the GBW model of the color dipole cross-section the higher twist terms are power-enhanced with decreasing $x$ (besides the generic suppression by negative powers of the hard scale), at twist $\tau$ one has $T_i ^{(\tau)} / T_i ^{(2)} \sim 1/(\mu^2 x^{\lambda})^{\tau-2}$ (modulo logarithms), with $\lambda \simeq 0.3$. Also the $q_T$-integrated structure functions are derived. Interestingly enough, this integration leads to the emergence of a twist~3 contribution in the $LT$~helicity structure function. Results of this paper provide the necessary tools for a forthcoming experiment oriented analysis of the LHC potential to measure the higher twist components of the proton structure within the color dipole approach, and prove that the applied twist decomposition method is effective.

\section{Kinematics and notation}

We consider the high energy proton--proton collision with a lepton--anti-lepton pair, $\lc^+ \lc^- = e^+ e^-$ or $\mu^+ \mu^-$, in the final state, $p(P_1)p(P_2) \to \lc^+ \lc^- X$ in which the pair is produced in the fragmentation region of one of the protons and the leptons four-momenta $l^+$ and $l^-$ are measured. At the leading order in QED this process is mediated by a virtual photon $\gamma^* (q)$ , with the four-momentum $q = l^+ + l^-$, and the virtuality $q^2 = M^2 > 0$ is the lepton pair invariant mass squared. It is convenient to introduce also $\kappa = l^+ - l^-$. The proton projectiles four momenta are $P_1$ and $P_2$ and they are near light-like, in the center of mass system (c.m.s.) of the $pp$ pair  $P_1 \simeq (\sqrt{\sp}/2,0,0,-\sqrt{\sp}/2)$, $P_2 \simeq (\sqrt{\sp}/2,0,0,\sqrt{\sp}/2)$, where the invariant collision energy squared $\sp=(P_1+P_2)^2$ is much greater than the proton mass squared, $m_p ^2$. We define light-like components of the momenta as $p^{\pm} = p^0 \pm p^z$, where the $z$ axis is given by the beam direction in the c.m.s. From now on we shall use the light-cone coordinates for four-vectors, $v=(v^+,v^-;\vec{v}_T)$. The Sudakov decomposition of four momenta will be employed, $p_i = \alpha_i P_1 + \beta_i P_2 + p_\perp$, where $p_\perp$ is the $p$ four-momentum component in the plane perpendicular to $P_1$ and $P_2$. For the DY virtual photon we have $q = \alpha_q P_1 + \beta_q P_2 + q_{\perp}$, with $q_{\perp} = (0,0;\vec{q}_T)$ and $q^2 _{\perp} = -\vec{q}_T^{\ 2}$. The forward region is defined by the condition $\beta \gg \alpha$, and the photon (or the DY pair) rapidity $y = 1/2 \log(\beta/\alpha)$. At the LHC the backward region $\alpha \gg \beta$ is, in fact, fully symmetric to the forward region, so for simplicity we shall restrict our discussion to  $\beta \gg \alpha$.

\begin{figure}

\centering
\includegraphics[width=.46\textwidth]{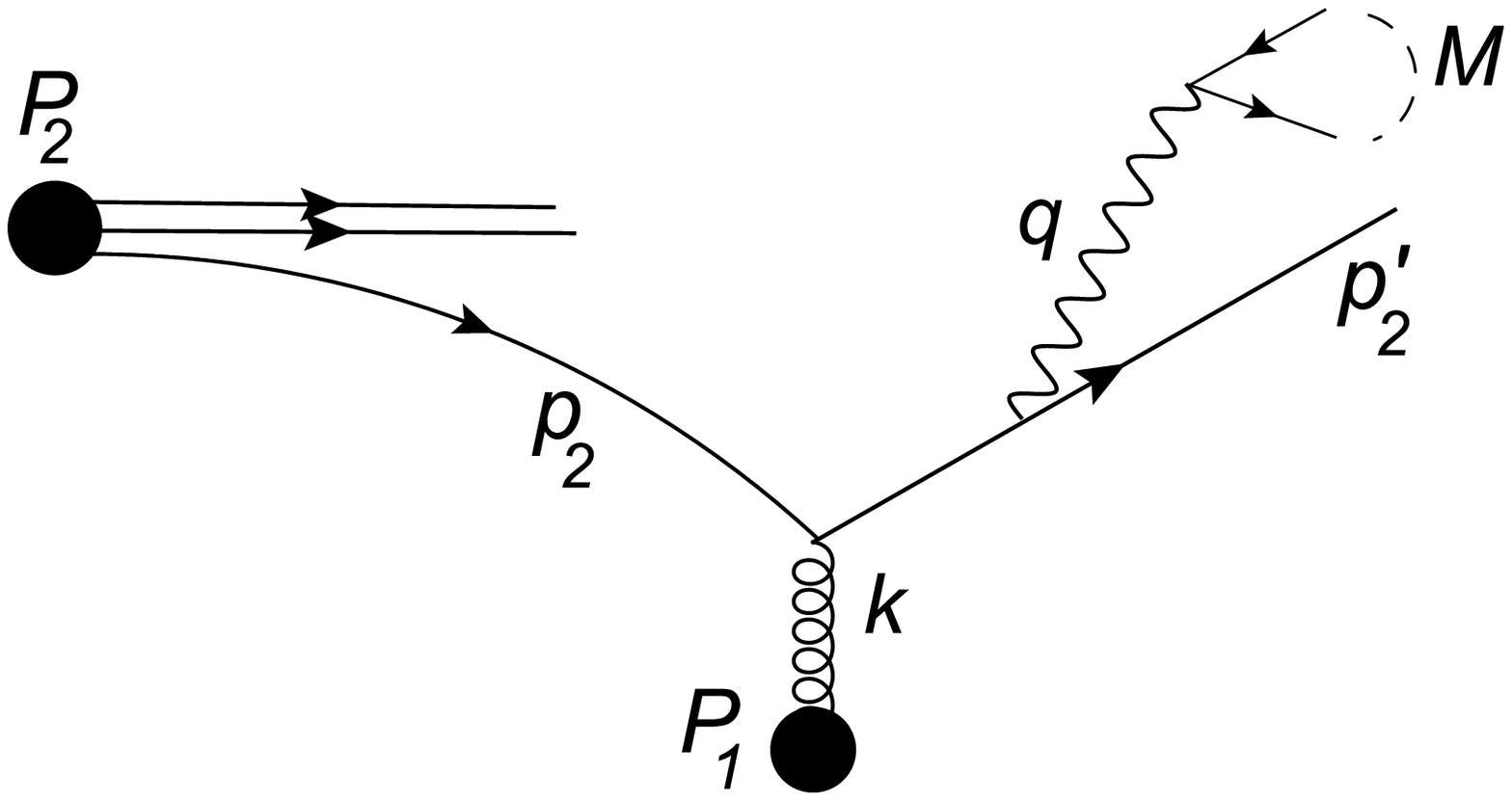}
\hfill
\includegraphics[width=.46\textwidth]{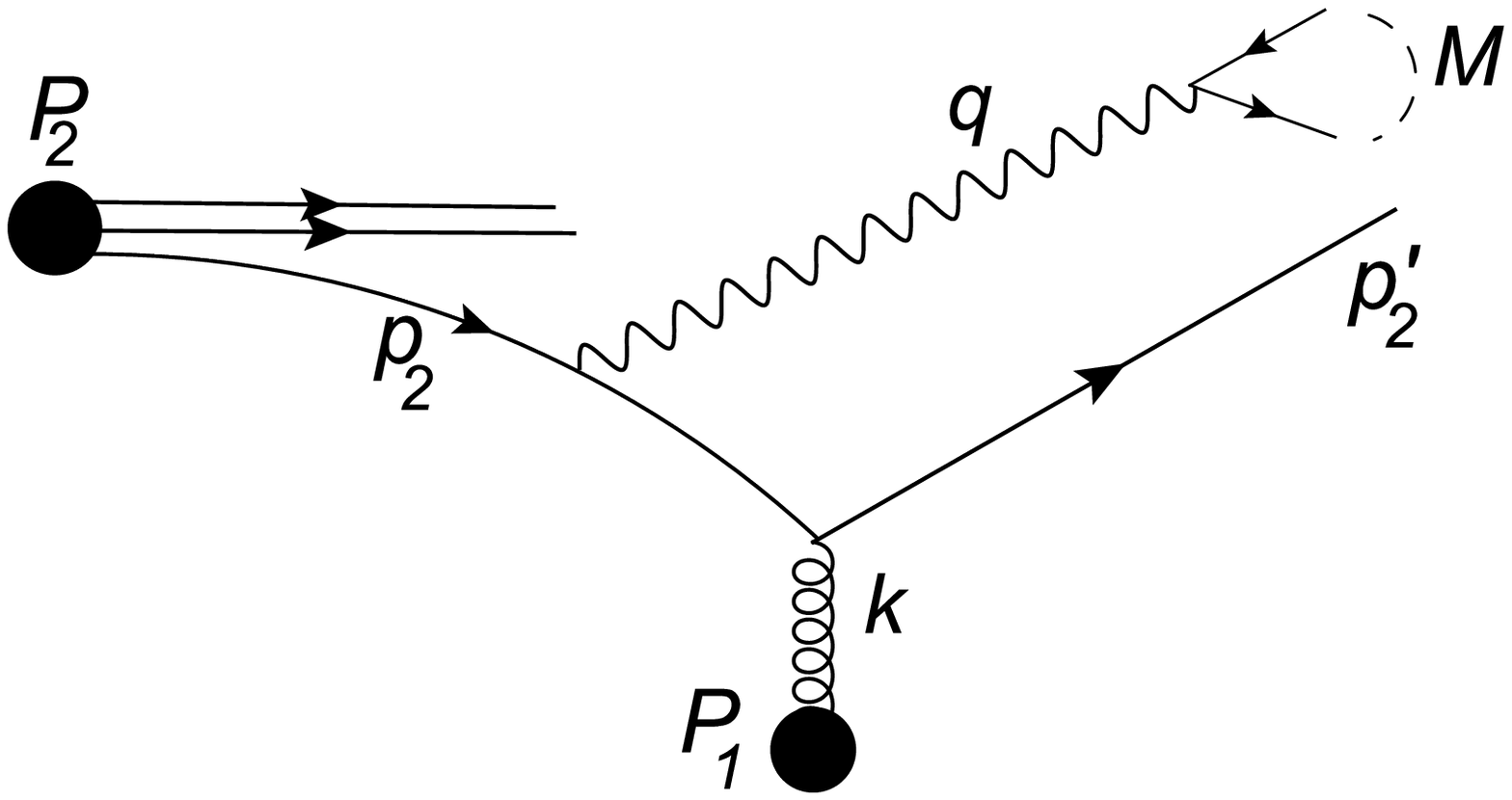}
\caption{Dominant diagrams for the forward Drell-Yan process in the $t$-channel helicity frame (in which the target is at rest), see the text for the notation. 
}
\label{leadingDiagrams}
\end{figure}

At the leading order of QED and QCD at the parton level the DY hard subprocess is $q_1(p_1)\bar q_2(p_2) \to \gamma^*(q) \to \lc^+ \lc^-$, where $q_1$ ($\bar q_2$) is a quark (anti-quark) coming from one of the protons (the other proton) that carries four momentum $p_1$ ($p_2$). At the NLO QCD the partonic subprocesses include also quark (anti-quark)--gluon contributions, $q_2(p_2) g(k) \to q_2 (p_2') \gamma^*(q) \to q_2 \lc^+ \lc^-$. In the high energy limit $p_1 = x_1 P_1 + p_{1 \perp}$ and $p_2 = x_2 P_2 + p_{2 \perp}$, where $x_i$ are the parton $x$-variables, see Fig.~\ref{leadingDiagrams}.

In the forward region at the LHC one has $x_2 \gg x_1$,  $x_2$ is a sizable fraction of the longitudinal proton momentum $P_2 ^+$, say $x_2 \sim 0.1$ and $x_1$ is very small, down to $x_1 \sim 10^{-6}$. In this region the quark $q_1$ distribution function (d.f.) is strongly dominated by the sea-quarks and the valence quark contribution to $q_1$ d.f.\ may be neglected. At a small~$x$ the evolution of the sea-quark (or anti-quark) distribution function of the target proton is driven by the gluon evolution. In more detail, due to the suppression of quark propagation over a large rapidity distance, the dominant contribution to the sea-quark distribution function comes from the diagrams where the sea-quark emerges from the gluon in the last splitting of the QCD evolution. Therefore, in the $k_T$ factorization framework the dominant diagrams with the gluon exchange in the $t$-channel are given in Fig.~\ref{leadingDiagrams}.
The topology of the diagram at the right-hand side coincides with the topology of the LO collinear contribution with the sea-quark emerging from the gluon at the last splitting. The diagram at the left-hand side of Fig.~\ref{leadingDiagrams} contributes to the NLO $qg$ partonic subprocess in the collinear picture.

The diagrams shown in Fig.~\ref{leadingDiagrams} are the basis of the color dipole approach to the forward DY process. Thus, one considers the virtual photon emission by a fast quark coming from $p(P_2)$ in the scattering off the proton $p(P_1)$ by a small-$x$ gluon exchange, $q(p_2)g(k) \to q(p'_2) \gamma^*(q)$. The effective color dipole emerges here through the interference of amplitudes of the virtual photon emission before and after the quark scattering off the target proton. The effective color dipole size corresponds to a displacement of the quark position in the transverse space due to the $\gamma^*$ emission.   
In the color dipole formulation one relies on the $k_T$-factorization (the high-energy factorization) approach, where the gluon transverse momentum does not vanish and the standard gluon distribution function $xg(x,\mu^2)$ gets replaced by a $k_T$ dependent unintegrated gluon distribution, $f_g(x, k_T^2)$ (with implicit scale dependence). At the LO, $xg(x,\mu^2) = \int^{\mu^2} dk_T^2 / k_T^2 \, f_g(x,k_T^2)$. On the other hand, the fast quark $q(p_2)$ carries a rather large $x_2$, so its transverse momentum may be neglected, $p_2$ is saturated by~$p_2 ^+$.

Having defined the relevant partonic channel and diagrams, we complete fixing the notation: let the gluon-$x$ be $x_g=k^-/P_1^-$, $z= q^+/p_2^+$ is the longitudinal momentum fraction of the fast quark $q(p_2)$ taken by $\gamma^*$ , $x_F=q^+/P_2^+$ is the Feynman $x$ of the virtual photon (or the DY pair). The fast quark distribution function in the proton $p(P_2)$, taken in the collinear limit, is denoted by $\wp$. Finally, it is convenient to analyze the process in the helicity basis, so we denote the helicities of the incoming and outgoing quark ($q(p_2)$ and $q(p'_2)$) by $\lambda_1$ and $\lambda_2$ respectively, and polarizations of $\gamma^*$ by $\sigma$. Since the photon is virtual, it has three polarization states.

\section{The forward Drell-Yan cross-section}
\subsection{Structure functions}

The standard description of the differential DY cross-section employs so-called helicity structure functions $W_L, W_T, W_{TT}, W_{LT}$ \cite{LamTung1,LamTung2} (see Appendix \ref{WvsT}). In this approach one factorizes leptonic and hadronic degrees of freedom by contracting both hadronic and leptonic tensors with virtual photon polarization vectors (PPVs). The leptonic tensor reduces to a distribution of lepton angles $\Omega=(\theta,\phi)$ in the lepton pair center-of-mass frame while the result of contraction of the hadronic tensor with the different PPVs are the $W$-structure functions. The differential DY cross-section is then given by the formula:
\begin{eqnarray}
\frac{d\sigma}{d x_F dM^2 d \Omega d^2 q_T} & = & \frac{\alpha^2_{\mathrm{em}}}{2(2\pi)^4 M^4} \left[ (1-\cos ^2 \theta) W_L + (1+\cos ^2 \theta) W_T + (\sin^2\theta \cos 2\phi)W_{TT} \right. \nonumber\\
& + & \left. (\sin2\theta \cos \phi) W_{LT}\right].
\label{sigAsWcomb}
\end{eqnarray}
The form of $W$-structure functions depends on an arbitrary choice of axes (which defines the PPVs) in the lepton pair center-of-mass frame. In this paper we perform calculations in a frame with the $Z$~axis anti-parallel to the target's momentum and the $Y$ axis orthogonal to the reaction plane (in \cite{LamTung1} this frame is called the $t$-channel helicity frame).

In order to avoid the helicity frame dependence one introduces invariant structure functions, $T_i$. They are defined as coefficients of the hadron tensor decomposition \cite{LamTung1}:
\begin{eqnarray}
W^{\mu \nu}= -T_1\ \tilde{g}^{\mu \nu}+T_2\  \tilde{P}^{\mu}  \tilde{P}^{\nu} - T_3\ \frac{1}{2}\left( \tilde{P}^{\mu}  \tilde{p}^{\nu}+\tilde{p}^{\mu}  \tilde{P}^{\nu}\right) +T_4\ \tilde{p}^{\mu}  \tilde{p}^{\nu}
\label{invDef0}
\end{eqnarray}
where $\tilde{g}^{\mu \nu}=g^{\mu \nu}-q^{\mu} q^{\nu}/q^2$, $P=P_1+P_2$, $p=P_1-P_2$ and $\tilde{P}^{\mu}=\tilde{g}^{\mu \nu} P_{\nu}/\sqrt{\sp}$, $\tilde{p}^{\mu}=\tilde{g}^{\mu \nu} p_{\nu}/\sqrt{\sp}$.
The invariant DY structure functions are related to the helicity structure functions in the $t$-channel helicity frame in the following way,
\begin{eqnarray}
T_1&=&W_T+W_{TT},\\ \nonumber
T_2&=&\frac{M^2}{x_F^2 \sp} W_L-\frac{M^2}{x_F^2 \sp} W_T-\frac{(M^2+\sp x_F^2)^2-2\sp x_F^2 q_T^2+q_T^4}{2 x_F^2 \sp q_T^2}W_{TT}+\frac{M(M^2+\sp x_F^2-q_T^2)}{x_F^2 \sp q_T}W_{LT}, \\  \nonumber
T_3&=&-\frac{2M^2}{x_F^2 \sp} W_L+\frac{2M^2}{x_F^2 \sp} W_T
-\frac{M^4-\sp^2 x_F^4+ q_T^4}{ x_F^2 \sp q_T^2}W_{TT}
+\frac{2M(-M^2+q_T^2)}{x_F^2 \sp q_T}W_{LT}, \\ \nonumber
T_4&=&\frac{M^2}{x_F^2 \sp} W_L-\frac{M^2}{x_F^2 \sp} W_T-\frac{(M^2-\sp x_F^2)^2+2\sp x_F^2 q_T^2+q_T^4}{2 x_F^2 \sp q_T^2}W_{TT}+\frac{M(M^2-\sp x_F^2-q_T^2)}{x_F^2 \sp q_T}W_{LT},
\label{matrixl}
\end{eqnarray}
see Appendix~\ref{WvsT} for the derivation. The DY helicity structure functions in any helicity frame may be expressed through the invariant structure functions and the explicit formulae for several standard frames may be found e.g.\ in Ref.~\cite{LamTung1}.

In the following we shall re-derive the DY helicity structure functions in the $k_T$-factorization approach. Thus, the scattering amplitudes of the fast quark will be computed within the high energy limit of QCD and the corresponding helicity dependent cross-sections will be represented in terms of the impact factors. In order to account for the multiple scattering effects we shall introduce the color dipole cross-section. Next the Mellin representations of the impact factors and of the helicity structure functions will be given.

\subsection{The forward Drell-Yan impact factors}

In the framework applied the forward DY cross-section takes the following form \cite{Brodsky,Kopeliovich,GJ},
\begin{eqnarray}
\frac{d\sigma}{d x_F dM^2 d \Omega d^2 q_T}&=&\frac{\alpha_{\mathrm{em}}}{(2\pi)^2(P_1\cdot P_2)^2 \ M^2 \ x_F^2(1-z)}  L^{\sigma \sigma'}(\Omega) \int_{x_F}^1 dz \ \wp(x_F/z) \nonumber \\
& & \times
\int d^2 k_T \frac{2\pi \alpha_s}{3} \frac{f({x}_g,k_T^2)}{k_T^4} \tilde{\Phi}_{\sigma \sigma'} (q_T,k_T,z),
\label{dsigma2}
\end{eqnarray}
where the  helicity dependent $\gamma^*$ impact factors are
\begin{eqnarray}
\tilde{\Phi}_{\sigma \sigma'} (q_T,k_T,z)= \sum_{\lambda_1,\lambda_2=+,-} \ A_{\lambda_1,\lambda_2}^{(\sigma)}(\vec{q}_T)^\dagger A_{\lambda_1,\lambda_2}^{(\sigma')}(\vec{q}_T),
\label{formfactor}
\end{eqnarray}
the leptonic tensor in the helicity basis reads
\begin{equation}
L^{\sigma \sigma'}=\epsilon^{(\sigma)}_{\mu}L^{\mu \nu} \ \epsilon_{\nu} ^{(\sigma')\dagger}, \ \ \  L^{\mu \nu}=-g^{\mu \nu}+\frac{\kappa^{\mu} \kappa^{\nu}}{\kappa^2},
\end{equation}
and $\wp(x_F/z)$ is a collinear parton distribution function for the projectile. The amplitudes $A_{\lambda_1,\lambda_2}^{(\sigma)}(\vec{q}_T)$ of the virtual photon emission with the polarization $\sigma$ and the transverse momentum $\vec{q}_T$ are given in Fig.~\ref{leadingDiagrams}.
In the target rest frame $P_1=(m_p,m_p;\vec{0})$ and we choose a standard set of the polarization vectors,
\begin{equation}
\epsilon^{(0)}=\left( \frac{q^+}{M},-\frac{M}{q^+};\vec{0} \right) \ \textrm{and} \ \  \epsilon^{(\pm)}=\left(0,0;\vec{\epsilon}\,{}_T ^{(\pm)} \right), \vec{\epsilon}\,{}_T^{(\pm)}=\frac{1}{\sqrt{2}}\left(1,\pm i \right).
\end{equation}

In the lepton c.m.s.\ where $\vec{l}^+=-\vec{l}^-$ we define the spatial axes $(X,Y,Z)$ through the $\gamma^*$ polarization vectors,
\begin{equation}
\epsilon^{(0)}_{\mu}=Z_{\mu} \ \textrm{and} \ \  \epsilon^{(\pm)}_{\mu}=\frac{1}{\sqrt{2}}\left( X_{\mu} \pm i Y_{\mu}  \right).
\end{equation}
The leptonic helicity tensors, $L^{\sigma \sigma'}$ are then expressed
through a set of three scalar products (recall that $\kappa = l^+ - l^-$),
\begin{eqnarray}
\kappa \cdot X&=&-2 |\vec{l}^+| \sin\theta \cos\phi,\\
\kappa \cdot Y&=&-2 |\vec{l}^+| \sin\theta\sin\phi, \\
\kappa \cdot Z&=&-2 |\vec{l}^+| \cos\theta,
\label{anglesDef}
\end{eqnarray}
where $\Omega=(\theta,\phi)$ are the standard angles of the spherical
coordinate system in the lepton CM frame.

With the chosen set of $\gamma^*$ polarization vectors the DY $\gamma^*$ emission amplitudes take the form
\begin{eqnarray}
A_{\lambda_1,\lambda_2}^{(0)}(\vec{q}_T) &=& \frac{e}{2\sqrt{\pi}(2\pi)^2}  \frac{\sqrt{1-z}}{z} \frac{x_F (P_1 \cdot P_2)}{M} \delta_{\lambda_1,\lambda_2} \nonumber
\\
& & \times \left[ \frac{M^2(1-z)}{M^2(1-z)+ \vec{q}_T^{\ 2}} - \frac{M^2(1-z)}{M^2(1-z)+ (\vec{q}_T-z\vec{k}_T)^2 }  \right]
\label{amplitudesMom0} ,
\end{eqnarray}
\begin{eqnarray}
A_{\lambda_1,\lambda_2}^{(\pm)}(\vec{q}_T) &=& \frac{e}{4\sqrt{\pi}(2\pi)^2} \frac{\sqrt{1-z}}{z} x_F (P_1 \cdot P_2)
\delta_{\lambda_1,\lambda_2}(2-z \mp \lambda_1 z) \nonumber \\
& & \times \left[ \frac{- \vec{q}_T }{M^2(1-z)+\vec{q}_T^{\ 2}} - \frac{-(\vec{q}_T-z\vec{k}_T) }{M^2(1-z)+(\vec{q}_T-z\vec{k}_T)^2 }  \right] \cdot \vec{\epsilon}_\bot^{\ (\pm)} ,
\label{amplitudesMom+}
\end{eqnarray}
where we suppressed the dependence of $A_{\lambda_1,\lambda_2}^{(\pm)}(\vec{q}_T)$ on $z$ and $\vec{k}_T$. The inverse Fourier transforms of the amplitudes to the transverse position space read
\begin{equation}
\tilde{A}_{\lambda_1,\lambda_2}^{(\sigma)}(\vec{r})
=\frac{1}{2 \pi} \int A_{\lambda_1,\lambda_2}^{(\sigma)}(\vec{q}_T) \  e^{-i \vec{q}_T \cdot \vec{r}}  \  d^2 q_T,
\label{FourierA}
\end{equation}
which is convenient to rewrite as
\begin{equation}
\tilde{A}_{\lambda_1,\lambda_2}^{(\sigma)}(\vec{r})=\left[ 1- e^{-i  z  \vec{k}_{T}\cdot \vec{r}} \right] \tilde{a}_{\lambda_1,\lambda_2}^{(\sigma)}(\vec{r}),
\label{smalla}
\end{equation}
where
\begin{eqnarray}
\tilde{a}_{\lambda_1,\lambda_2}^{(0)}(\vec{r}) &=& \frac{e}{2\sqrt{\pi}(2\pi)^2}  \frac{\sqrt{1-z}}{z} x_F (P_1 \cdot P_2) \delta_{\lambda_1,\lambda_2} M (1-z) K_0 \left(  \sqrt{1-z} M r \right),\\
\tilde{a}_{\lambda_1,\lambda_2}^{(\pm)}(\vec{r}) &=& i \frac{e}{4\sqrt{\pi}(2\pi)^2} \frac{\sqrt{1-z}}{z} x_F (P_1 \cdot P_2)
\delta_{\lambda_1,\lambda_2} \nonumber \\
& & \times (2-z \mp \lambda_1 z) M \sqrt{1-z}
 K_1 \left( \sqrt{1-z} M r \right) \frac{r_x\pm ir_y}{\sqrt{2} \ r}.
\end{eqnarray}

Using representation (\ref{smalla}) of the amplitudes, with the dipole eikonal factor, $1-e^{-iz\vec{k}\cdot\vec{r}}$, one may express the forward DY cross-section (\ref{dsigma2}) in terms of the color dipole cross-section \cite{Brodsky,Kopeliovich},
\begin{equation}
\hat{\sigma}(r)=\frac{2\pi \alpha_s}{3}  \int d^2 k_T \frac{ f(\bar{x}_g,k_T^2) }{k_T^4} \ \big|1- e^{-i \vec{k}_T\cdot \vec{r}} \big|^2,
\end{equation}
where we introduce $\bar x_g$ being the value of $x_g$ at the process threshold.  In what follows we suppress the $\bar x_g$ dependence of $\hat{\sigma}(r)$. At the leading order the dipole cross-section is in one-to-one correspondence with the unintegrated gluon density,
$f(\bar{x}_g,k_T^2)$, and their inverse relation reads
\begin{equation}
\frac{2\pi \alpha_s}{3} \frac{f(\bar{x}_g,k_T^2)}{k_T^2}=\frac{1}{2} \int d^2r \ e^{i \vec{k}_{T}\cdot \vec{r}} \nabla^2 \hat{\sigma}(r)= \frac{1}{2} \int d^2r \ e^{i \vec{k}_{T}\cdot \vec{r}} \ \hat{\sigma}(r) (-\vec{k}_T^{\ 2}),
\end{equation}
where $\nabla^2$ is the Laplace operator in two transverse dimensions.

Using the last formula one can rewrite (\ref{dsigma2}) as
\begin{eqnarray}
\frac{d\sigma}{d x_F dM^2 d \Omega d^2 q_T}&=&\frac{\alpha_{\mathrm{em}}}{(2\pi)^2(P_1\cdot P_2)^2 \ M^2 \ x_F^2(1-z)}  L^{\sigma \sigma'}(\Omega) \nonumber \\
& & \times \int_{x_F}^1 dz \ \wp(x_F/z)
\int d^2 r \  \hat{\sigma}(r) \Phi_{\sigma \sigma'} (q_T,r,z)
\label{dsigma3}
\end{eqnarray}
with
\begin{eqnarray}
\Phi_{\sigma \sigma'} (q_T,r,z)= -\frac{1}{2} \int d^2 k_T \ e^{i \vec{k}_{T}\cdot \vec{r}} \tilde{\Phi}_{\sigma \sigma'} (q_T,k_T,z).
\label{fourierTrPhi}
\end{eqnarray}

Substituting the inverse Fourier transform of (\ref{FourierA}) into (\ref{fourierTrPhi}) and integrating over  $d^2 k_T$ one obtains the impact factors in the transverse position representation,
\begin{eqnarray}
\Phi_{\sigma \sigma'} (q_T,r,z)=\frac{1}{2}
\sum_{\lambda_1,\lambda_2=+,-} \int d^2 r_1 d^2 r_2 \ \tilde{a}_{\lambda_1,\lambda_2}^{(\sigma)}(\vec{r}_1)^\dagger \tilde{a}_{\lambda_1,\lambda_2}^{(\sigma')}(\vec{r}_2) \  e^{-i \vec{q}_T \cdot (\vec{r}_1-\vec{r}_2)}  \nonumber
\\
 \times
\bigg[  \delta(\vec{r}-\vec{r_1}) +  \delta(\vec{r}-\vec{r_2}) - \delta(\vec{r}-(\vec{r_1}-{r_2}))    \bigg].
\label{formfactor3}
\end{eqnarray}

Finally, let us parameterize the DY structure functions $W_i$ in terms of new functions~$\Phi_i$,
\begin{eqnarray}
 W_{i}=\frac{2(2\pi)^4 M^4}{\alpha^2_{\mathrm{em}}}
 \int_{x_F}^1 dz \ \wp(x_F/z)
\int d^2 r \  \hat{\sigma}(r)  \Phi_{i} (q_T,r,z)
\label{Wphi}
\end{eqnarray}
for $i=\left\{ L,T,TT,LT \right \}$.
Comparing (\ref{sigAsWcomb}) and (\ref{dsigma3}) one gets the following expressions for $\Phi_i$ in terms of the DY $\gamma^*$ impact factors :
\begin{eqnarray}
 L^{00}(\Omega) \Phi_{00} (q_T,r,z)
& \equiv &
(1-\cos ^2 \theta)\Phi_L(q_T,r,z),
\label{PhiPhiL} \\
%
 L^{++}(\Omega) \Phi_{++} (q_T,r,z)+L^{--}(\Omega) \Phi_{--} (q_T,r,z) 
& \equiv &
(1+\cos ^2 \theta) \Phi_T(q_T,r,z), \\
%
 L^{+-}(\Omega) \Phi_{+-} (q_T,r,z)+L^{-+}(\Omega) \Phi_{-+} (q_T,r,z) 
& \equiv &
 (\sin^2\theta \cos 2\phi) \Phi_{TT}(q_T,r,z), \\
%
 L^{0+}(\Omega) \Phi_{0+} (q_T,r,z)+L^{0-}(\Omega) \Phi_{0-} (q_T,r,z) 
 & & \nonumber \\
+L^{+0}(\Omega) \Phi_{+0} (q_T,r,z)+ L^{-0}(\Omega) \Phi_{-0} (q_T,r,z) 
 & \equiv &
(\sin2\theta \cos \phi) \Phi_{LT}(q_T,r,z).
\label{PhiPhiLT}
\end{eqnarray}
The functions $\Phi_{\sigma\sigma'}$ and $\Phi_i$ play a similar r\^{o}le to the virtual photon wave functions of the color dipole model for the DIS, but for the DY also off-diagonal amplitude products in the helicity basis contribute. The functions $\Phi_{\sigma\sigma'}$ are related by (\ref{fourierTrPhi}) to the standard impact factors in the momentum space, so we shall use for them the term impact factors as well, specifying the $\Phi_i$ functions as the {\em leptonic impact factors}.

\subsection{The Mellin representation of the impact factors}

Mellin representation is a useful tool in the analysis of QCD amplitudes, in particular of their twist structure. In order to find this representation for the forward DY process we start introducing a Mellin transform of the dipole cross-section,
\begin{equation}
 \hat{\sigma}(\vec{r})=\int_{\cal C} \frac{ds}{2\pi i} \left ( \frac{ Q_0^2}{4} r \right ) ^s \tilde{\sigma} (-s),
\label{invMellin}
\end{equation}
where the contour ${\cal C}$ is a vertical line in the complex~$s$ plane, $Q_0$ is the saturation scale and
\begin{equation}
\tilde{\sigma}(-s)=\int_0^\infty \frac{d \rho^2}{\rho^2} \left(\rho^2 \right)^{-s} \
 \hat{\sigma}(\vec{\rho}) .
\label{Mellin}
\end{equation}
Using (\ref{invMellin}) we rewrite (\ref{Wphi}) as:
\begin{eqnarray}
 W_{i}=\int_{x_F}^1 dz \ \wp(x_F/z)
\int_{\cal C} \frac{ds}{2\pi i} \ \tilde{\sigma} (-s) \left ( \frac{z^2 Q_0^2}{\eta^2_z} \right) ^s \hat{\Phi}_{i} (q_T,s,z),
\label{Wphihat}
\end{eqnarray}
where $\eta^2_z=M^2 (1-z)$ and
\begin{eqnarray}
 \hat{\Phi}_{i} (q_T,s,z)=\frac{2(2\pi)^4 M^4}{\alpha^2 _{\mathrm{em}}} \int d^2 r \   \left ( \frac{\eta^2_z}{4z^2}\ r \right) ^s  \Phi_{i} (q_T,r,z)
\label{PhiMellin}
\end{eqnarray}
is the Mellin representation of the leptonic impact factor.

The leptonic impact factors $\Phi_{i} (q_T,r,z)$ are linear combinations of  $\Phi_{\sigma \sigma'} (q_T,r,z)$ (see formulae (\ref{PhiPhiL})---(\ref{PhiPhiLT})) where $\Phi_{\sigma \sigma'}$ are given in (\ref{formfactor3}).
After integration (\ref{PhiMellin}) over  $d^2 r$ and  $d^2 r_1d^2 r_2$ we get the following results for Mellin transforms of the leptonic impact factors,
\begin{eqnarray}
\hat{\Phi}_{L} (q_T,s,z)&=& \frac{2}{z^2} \left\{ \frac{2 \Gamma^2(s+1) }{1+q_T^2/\eta^2_z} \ {}_2 F_1 \left (s+1,s+1,1,-\frac{q_T^2}{\eta^2_z} \right)
\right. \nonumber \\
& & - \left.
\Gamma(s+1) \Gamma(s+2) \ {}_2 F_1 \left (s+1,s+2,1,-\frac{q_T^2}{\eta^2_z} \right) \right\} ,
\label{sigmaLnonInt} 
\end{eqnarray}
\begin{eqnarray}
\hat{\Phi}_{T} (q_T,s,z) &=&
\frac{1+(1-z)^2}{2z^2(1-z)}\Bigg\{
\frac{2 q_T^2/\eta^2_z}{1+q_T^2/\eta^2_z} \Gamma(s+1) \Gamma(s+2) \ {}_2 F_1 \left(s+1,s+2,2,-\frac{q_T^2}{\eta^2_z} \right) \Bigg. \nonumber \\
& & - \Gamma(s+1)^2
\left[
{}_2 F_1 \left(s+1,s+1,1,-\frac{q_T^2}{\eta^2_z} \right)  \nonumber
\right.
\\
& & \Bigg.  \left.
-(s+1)\ {}_2 F_1 \left (s+1,s+2,1,-\frac{q_T^2}{\eta^2_z} \right) \right] \Bigg\},
\label{sigmaTnonInt} \\
\hat{\Phi}_{TT} (q_T,s,z) &=&  \frac{1}{2 z^2}\left\{ \frac{2\pi }{\Gamma(1-s)\sin\pi s \ q_T^2/\eta^2_z} \left(1+\frac{q_T^2}{\eta^2_z} \right)^{-s-3} \Gamma(s+2)
\nonumber \right.\\
& & \left. \left[ \left(1+\frac{q_T^2}{\eta^2_z} \right)\left(1+\frac{q_T^2}{\eta^2_z}(s+2) \right)  \ {}_2 F_1 \left (-s+1,s+1,1,\frac{q_T^2}{q_T^2+\eta^2_z} \right) \right. \right. \nonumber \\
& & \left.  - \left(1+2\frac{q_T^2}{\eta^2_z}(s+1) \right)  {}_2 F_1 \left (-s+1,s+2,1,\frac{q_T^2}{q_T^2+\eta^2_z} \right) \right]\nonumber \\
& & - \left. \frac{4q_T^2/\eta^2_z}{1+q_T^2/\eta^2_z} \Gamma(s+1) \Gamma(s+2) \ {}_2 F_1 \left (s+1,s+2,2,-\frac{q_T^2}{\eta^2_z} \right)  \right\} ,
\label{sigmaTTnonInt}\\
\hat{\Phi}_{LT} (q_T,s,z)&=&\frac{2-z}{z^2 \sqrt{1-z}} \left\{  \pi \frac{q_T/\eta_z}{(1+q_T^2/\eta^2_z)^{s+2}} \ \frac{\Gamma(s+2)}{\Gamma(-s-1)\sin\pi s} \ {}_2 F_1 \left (-s,s+2,2,\frac{q_T^2}{q_T^2+\eta^2_z} \right) \right.
 \nonumber \\
& & - \left. \frac{q_T/\eta_z}{1+q_T^2/\eta^2_z} \Gamma^2(s+1)\ \left[ \  {}_2 F_1 \left (s+1,s+1,1,-\frac{q_T^2}{\eta^2_z} \right) \right. \right. \nonumber \\
& & + \left. \left. (s+1) \  {}_2 F_1 \left (s+1,s+2,2,-\frac{q_T^2}{\eta^2_z} \right) \right] \right\} .
\label{sigmaLTnonInt}
\end{eqnarray}
The above formulae are one of the main results of this paper. Inserted into (\ref{Wphihat}) they allow for an efficient analysis of the forward DY cross-section in terms of the double integrals over $z$ and $s$. Let us stress that the obtained Mellin forms of the impact factors do not depend on the chosen form of the color dipole cross-section, they follow directly from the perturbative (leading order) QCD amplitudes in the high energy limit.

\subsection{The Mellin representation of the integrated helicity structure functions}

For a description of inclusive measurements of the DY processes one applies $q_T$-integrated helicity structure functions,
\begin{equation}
\tilde{W}_i=\frac{1}{2\pi M^2} \int W_i \ d^2 q_T .
\end{equation}
Clearly, the results depend on the chosen set of the virtual photon polarization vectors, in particular one has to account carefully for a possible correlation of the polarization vectors with $\vec{q}_T$. $\tilde{W}_i$ can be found by the direct integration of the helicity impact factors $\hat{\Phi}_{i} (q_T,s,z)$ (given by (\ref{sigmaLnonInt})---(\ref{sigmaLTnonInt})) over  $d^2 q_T$ and then use (\ref{Wphihat}). In the $t$-channel helicity frame the integration over the azimuthal angle is trivial and gives $2 \pi$, and the remaining integrals can be performed using a new variable $q_T/ \eta_z$. One obtains:
\begin{eqnarray}
 \tilde{W}_{L} & = &\int_{\cal C} \frac{ds}{2\pi i}\ \int_{x_F}^1 dz \  \wp(x_F/z) \frac{1-z}{z^2}\ \left ( \frac{z^2 Q_0^2}{4 \eta^2_z} \right) ^s \tilde{\sigma}(-s)
 \left\{ \frac{\sqrt{\pi}\  \Gamma^3(s+1)}{\Gamma\left(s+\frac{3}{2}\right)}  \right\} ,
\label{WtyldeL} 
\end{eqnarray}
\begin{eqnarray}
\tilde{W}_{T}&=& \int_{\cal C} \frac{ds}{2\pi i}\  \int_{x_F}^1 dz \ \wp(x_F/z) \frac{1+(1-z)^2}{z^2}
\left ( \frac{z^2 Q_0^2}{4 \eta^2_z} \right) ^s \tilde{\sigma}(-s)
\nonumber \\
& &
\times \left\{
\frac{\sqrt{\pi}\ \Gamma(s) \Gamma(s+1) \Gamma(s+2)}{4\Gamma\left(s+\frac{3}{2}\right)}
\right\} ,
\label{WtyldeT} \\
\tilde{W}_{TT}&=& \int_{\cal C} \frac{ds}{2\pi i} \  \int_{x_F}^1 dz \  \wp(x_F/z)\frac{1-z}{z^2}
 \left ( \frac{z^2 Q_0^2}{4 \eta^2_z} \right) ^s \tilde{\sigma}(-s) \nonumber \\
& & \times
\left\{
\frac{\Gamma(s) \Gamma(s+1) \left[  4^s\  \Gamma\left(s+\frac{3}{2}\right)  -\sqrt{\pi}\ \Gamma(s+2)  \right]}{2\Gamma\left(s+\frac{3}{2}\right)}
\right\} ,
\label{WtyldeTT} \\
\tilde{W}_{LT}&=& \int_{\cal C} \frac{ds}{2\pi i} \ \int_{x_F}^1 dz \ \wp(x_F/z) \frac{(2-z)\sqrt{1-z}}{z^2}
 \left ( \frac{z^2 Q_0^2}{4 \eta^2_z} \right) ^s \tilde{\sigma}(-s) \nonumber \\
& & \times \left\{
-4^{s-1} \ \Gamma^2\left(s+\frac{1}{2} \right)+ \chi_1(s)+\chi_2(s)
\right\},
\label{WtyldeLT}
\end{eqnarray}
where functions $\chi_{1,2}$ are define by their integral representations
\begin{eqnarray}
\chi_1(s)&=&\int_0^\infty \frac{dt}{(1+t^2)^{1/2}} \int_0^\infty d\rho \ \rho^{2s+1} \sin(\rho t) K_0(\rho),
\nonumber \\
\chi_2(s)&=&-\int_0^\infty \frac{dt}{(1+t^2)^{3/2}} \int_0^\infty d\rho \ \rho^{2s} \sin(\rho t) K_1(\rho).
\label{kappaDef}
\end{eqnarray}
These expressions provide a compact representation of the $q_T$-inclusive forward DY cross-sections differential in the angles of DY leptons in the
$t$-channel helicity frame.

\section{The twist expansion of the DY structure functions and the Lam-Tung relation}

With the Mellin representations of the leptonic DY impact factors derived in the previous section, it is possible to evaluate the DY structure functions and perform their twist analysis. For this one needs, however, to specify the dipole cross-section, $\hat\sigma(\vec \rho)$. Within OPE approach to hard scattering in QCD, $\hat\sigma(\vec \rho)$ includes a tower of matrix elements with increasing twist. Currently the higher twist components of the proton structure at small~$x$ are not known from the experiment. Also a theoretical analysis within perturbative QCD is not capable to predict the non-perturbative inputs for evolution of the higher twist operators from the first principles. Therefore current theoretical estimates of the higher twist effects must rely on certain assumptions and models. So far two main QCD-inspired approaches to the dipole cross-sections beyond the leading twist were used in the higher-twist analysis: the eikonal Glauber-Mueller scattering cross-section (used in the GBW saturation model) and the Balitsky-Fadin-Kuraev-Lipatov (BFKL) and the Balitsky-Kovchegov (BK) dipole cross-sections, obtained within the small-$x$ resummation framework in QCD \cite{bfkl,bfklrev,Balitsky,Kov}. With both forms of the dipole cross-sections one can describe well the existing data on the proton structure from the deep inelastic scattering (DIS) at small~$x$, see e.g.\ Ref.\ \cite{KMW}, but as it was found in Refs.\ \cite{BGBM,MS}, the inclusive DIS data probe the proton structure efficiently only at the leading twist. Hence, the current theoretical predictions of the higher twist effects are uncertain. Still, it is important to estimate these effects within different models in order to design measurements probing the higher twist effects and to make an optimal use of the resulting data. In particular, measurements of the higher twist contributions in the forward DY structure functions may be used to discriminate between models of the higher twist / multiple scattering effects and provide essentially new information about the proton structure and the QCD evolution beyond the leading twist.

As a first step towards understanding of the higher twist structure of the forward DY scattering we consider the simplest case of the eikonal dipole cross-section used in the GBW approach. Using this cross-section we perform an explicit analytic twist decomposition of the forward DY structure functions. This test case exhibits already some interesting features, clearly visible due to a simple analytic form of the results. A more detailed numerical analysis of the forward DY process at the LHC, taking into account also the BFKL / BK dipole cross-section will be performed in a forthcoming paper \cite{MSS2}.

\subsection{The twist decomposition of the helicity structure functions $W_i$}

The twist decomposition of the forward DY structure functions $W_i$
may be performed using the Mellin representation (\ref{Wphihat}),
(\ref{sigmaLnonInt})---(\ref{sigmaLTnonInt}) supplemented by the GBW dipole cross section, that takes the form \cite{GBW},
\begin{equation}
\hat{\sigma}(\vec{\rho})=\sigma_0(1-e^{-\rho^2}),
\end{equation}
and its Mellin transform reads $\tilde{\sigma}(-s)=-\sigma_0 \Gamma(-s)$.

In order to define properly the twist content of the DY structure functions one should specify the scale that plays the r\^{o}le of the OPE hard scale. The forward DY structure functions (\ref{Wphihat}), however, depend on two hard scales, $M$ and $q_T$, as visible from the form of the impact factors (\ref{sigmaLnonInt})---(\ref{sigmaLTnonInt}), so the choice is ambiguous. In order to avoid this ambiguity we take advantage of the fact that in the OPE series the powers of the hard scale $\mu$ are accompanied by the corresponding opposite powers of the hadronic (soft) scale $\Lambda$, so that twist-$\tau$ terms enter with scale ratio $(\Lambda / \mu)^{\tau}$. In the color dipole formulation the soft scale enters only through the dipole cross-section in accordance with the requirement that the soft scale should be related to the target hadron structure. The GBW dipole cross-section carries a unique soft scale --- the saturation scale $Q_0(\bar x_g)$. In order to match the OPE series structure we identify $\Lambda = Q_0$ and define the twist-$\tau$ terms as those that scale in $Q_0$ as $Q_0 ^{\tau}$.\footnote{The na\"{i}ve scaling is modified by the QCD evolution leading to anomalous dimensions. This may be also taken into account in our approach to twist analysis as shown in \cite{BGBM}, but the anomalous dimensions vanish in the GBW model so the na\"{i}ve scaling of twist contributions holds (modulo logarithms).}.
The problem of the ambiguous hard scale may be understood from the OPE perspective. In the OPE the forward DY structure functions depend on non-perturbative matrix elements of hadronic operators and perturbative coefficient functions. The hadronic operators depend on the hadronic scale $\Lambda$ and the factorization scale $\mu$, and the coefficient functions depend on $M$, $q_T$, and $\mu$. Hence, if the factorization scale $\mu$ is identified with one of the hard scales, $M$ or $q_T$, then in the DY structure functions computed within the color dipole approach dependencies on $M$ and $q_T$ coming from the matrix element and from the coefficient functions are entangled and, in general, the scaling in $M$ or $q_T$ cannot be used to isolate the terms with definite twist. On the other hand, the procedure based on the determination of terms with the definite positive powers of $Q_0$ leads to a unique definition of the twist expansion in the adopted approach.

The twist decomposition of the DY structure functions reduces, therefore, to determination of their $Q_0$ power series expansion. This may be done with a technique described in Refs.\ \cite{BGBP,BGBM,GBLS}, based on analytic
structure in $s$ of the Mellin representation. One closes the contour ${\cal C}$ of the inverse Mellin transform  (\ref{Wphihat}) with a semicircle at complex infinity for $\mathrm{Re}\; s > 0$. The integrands are analytic in $s$ except of isolated singularities so one can express the inverse Mellin integrals as sums of contributions coming from the singularities in the complex~$s$ and these contributions carry a definite $Q_0$ scaling, hence the definite twist. For the GBW dipole cross-section the only singularities in the complex plane of the Mellin variable~$s$ in integrands (\ref{Wphihat}) are the simple poles coming from $\Gamma(-s)$, contained in the dipole cross-section, so the evaluation of the $s$-integrations by taking the residues is straightforward. Integrals over $z$ in (\ref{Wphihat}) are convergent for $\mathrm{Re}\ s>0$. The obtained leading twist contributions to the forward DY structure functions are the following,
\begin{eqnarray}
W_L^{(2)} &=& \s0 \frac{Q_0^2}{M^2} \int_{x_F}^1 dz \  \wp(x_F/z)  \frac{4M^{6}\ q_T^2 (1-z)^2}{\left[q_T^2+M^2(1-z) \right]^4} ,
\label{WLtw2} \\
W_T^{(2)} &=& \s0 \frac{Q_0^2}{M^2} \int_{x_F}^1 dz \  \wp(x_F/z)
\left[1+(1-z)^2\right] \frac{M^{4}\left[ q_T^4 +M^4 (1-z)^2\right]}{2\left[q_T^2+M^2(1-z) \right]^4} ,
\label{WTtw2} \\
W_{TT}^{(2)} &=& \s0 \frac{Q_0^2}{M^2} \int_{x_F}^1 dz \  \wp(x_F/z)  \frac{2M^{6}\ q_T^2 (1-z)^2}{\left[q_T^2+M^2(1-z) \right]^4} ,
\label{WTTtw2} \\
W_{LT}^{(2)}&=& \s0 \frac{Q_0^2}{M^2} \int_{x_F}^1 dz \ \wp(x_F/z) (2-z) \ \frac{M^{5}\ q_T \left[ -q_T^2 +M^2 (1-z)\right] (1-z)}{\left[q_T^2+M^2(1-z) \right]^4}\, .
\label{WLTtw2}
\end{eqnarray}
It is clearly visible from the above equations, that the structure functions have the following leading behavior at small values of the photon transverse momentum, $q_T \ll M$:  $W_T^{(2)} / \sigma_0 \sim O(1)$,  $W_{LT}^{(2)} / \sigma_0 \sim O(q_T/M)$, $W_{L,TT}^{(2)} / \sigma_0 \sim O(q_T^2/M^2)$. Therefore, there is a hierarchy  of contributions in which $W_T^{(2)}$ dominates over $W_{LT}^{(2)}$ and finally come $W_{L,TT}^{(2)}$.

At twist~4 one finds,
\begin{eqnarray}
W_L^{(4)} &=& \s0
 \frac{Q_0^4}{M^4} \int_{x_F}^1 dz \  \wp(x_F/z) z^2 \times \nonumber \\
 & & \times \frac{4 M^{8} \left[7 q_T^2 -10 M^2  q_T^2 (1-z) +
    M^4 (1-z)^2 \right] (1-z)^2}{\left[q_T^2+M^2 (1-z) \right]^6} ,
\label{sigmaLtw4}
\\
W_T^{(4)} &=& \s0
 \frac{Q_0^4}{M^4} \int_{x_F}^1 dz \  \wp(x_F/z)
\left[1+(1-z)^2\right]z^2 \times  \nonumber \\
 & & \times \frac{M^{6}\left[q_T^2 -2M^2 (1-z)\right]\left[q_T^4
  - 4 M^2 q_T^2 (1-z)+  M^4 (1-z)^2  \right]}{\left[q_T^2+M^2 (1-z) \right]^6} ,
\label{sigmaTtw4}\\
W_{TT}^{(4)} &=& \s0
 \frac{Q_0^4}{M^4} \int_{x_F}^1 dz \  \wp(x_F/z) z^2  \frac{12 M^{8}  q_T^2 \left[q_T^2 -2 M^2 (1-z) \right] (1-z)^2}{\left[q_T^2+M^2 (1-z) \right]^6} ,
\label{sigmaTTtw4}
\end{eqnarray}
\begin{eqnarray}
W_{LT}^{(4)}&=& \s0
\frac{Q_0^4}{M^4} \int_{x_F}^1 dz \ \wp(x_F/z) (2-z) \  z^2 \times
\nonumber \\
& & \times \frac{2M^{7}\ q_T \left[ -2 q_T^2 +M^2 (1-z) \right] \left[q_T^2 - 5M^2 (1-z) \right](1-z)}{\left[q_T^2+M^2 (1-z) \right]^6} .
\label{sigmaLTtw4}
\end{eqnarray}
In general, only the even twist contributions are found in the structure functions. Evaluation of twist $\tau >4$ contributions may be easily done as well.

The obtained results on the twist content of the structure functions exhibit certain common features. Clearly, at twist $\tau$ one has a suppression by the negative powers of the scale and an enhancement due to the growth of the saturation scale
$Q_0(\bar x_g) \sim \bar x_g ^{-\lambda}$ with the decreasing $\bar x_g$, with $\lambda \simeq 0.3$.
Hence $W_i ^{(\tau)} \sim 1/(M^2 \bar x_g ^{\lambda})^{\tau}$. This small $x_g$-enhancement  is specific to the eikonal color dipole model and not necessarily holds in other models. One sees as well that the two hard scales of the problem, $M$ and $q_T$, enter in combinations dependent on the term and none of them can be identified as the unique main hard scale of the problem. Moreover, the denominators of the integrands are proportional to powers of $\bar M^2 = M^2 (1-z) + q^2 _{\bot}$. Thus, in the region $z \to 1$ and $q_T \to 0$ the scale $\bar M \to 0$ and one finds soft singularities. We will discuss the treatment of this problem in Sec.\ 4.3.\


The terms of twist expansions of the helicity structure functions (\ref{WLtw2})---(\ref{WLTtw2}) and (\ref{sigmaLtw4})---(\ref{sigmaLTtw4}), are valid in the $t$-channel helicity frame only. However, these results may be used to obtain the twist expansion of the invariant structure functions, $T_i$. Twist-2 contributions to the invariant structure functions are given in Appendix~\ref{apTinv}.

\subsection{The twist expansion of the integrated structure functions}

It is easy to see that the expressions for the twist~2 and twist~4 components of the DY structure functions derived in the previous sections cannot be simply integrated over $ d^2 q_T$.  The reason is a divergence at $q_T \to 0$, $z\to 1$, that follows as a consequence of the form of denominators in the integrands,
$\propto (q_T ^2 + M^2(1-z))^n$. For example, the $q_T$ integration in $W_T ^{(2)}$ leads to a divergent $z$-integral:
\begin{eqnarray}
\int W_T^{(2)} \ d^2 q_T=\s0 \frac{Q_0^2}{M^2} \int_{x_F}^1 dz \  \wp(x_F/z)
\frac{1+(1-z)^2}{1-z} \frac{\pi M^2 }{3}\, .
\end{eqnarray}
In order to define properly the twist expansion of the integrated structure
functions one has to perform the $q_T$ integration of (\ref{WtyldeL})---(\ref{WtyldeLT}) prior to evaluating the inverse Mellin integral and carrying out the
twist decomposition. Then, the integration over $ d^2 q_T \, dz$ introduces additional poles at the positive integer values of the Mellin variable~$s$, and double and single poles in $s$ are found in the Mellin representation of the integrated DY structure functions. The problem with $z\to1$ limit of the integrands in the twist expansion of the integrated forward DY structure functions was found and solved in Ref.\ \cite{GBLS}. In short, in this prescription
one assumes a power series structure of the structure function integrands in the variable $1-z$  (modulo logarithms) at $z\to 1$ and treats analytically terms of this series that would na\"{i}vely lead to the divergent integrals of the twist components. The corresponding integrals lead to additional poles in $s$, $\sim 1/(s-\tau)$ at twist ~$\tau$. The $s$~integration of these terms may be then performed analytically. The additional $s$-singularity increases the order of the twist poles by one, which results with an additional logarithm $\ln M^2 / Q^2$ in the structure functions. The remaining part of the structure function integrands left after separation of the apparently singular terms, lead to the convergent integrals that can be directly evaluated by a numerical integration.

Thus, following the prescription of Ref.\ \cite{GBLS} we get for the leading twist,
\begin{eqnarray}
 \tilde{W}^{(2)}_{L}=  \s0 \frac{Q_0^2}{3 M^2}  \int_{x_F}^1 dz  \ \wp(x_F/z),\;\;
 \tilde{W}^{(2)}_{TT}=  \s0 \frac{Q_0^2}{6 M^2}  \int_{x_F}^1 dz  \ \wp(x_F/z),\;\; \tilde{W}^{(2)}_{LT}= 0,
\end{eqnarray}
\begin{eqnarray}
 \tilde{W}^{(2)}_{T}= \s0 \frac{Q_0^2}{4 M^2}\Bigg\{  \wp(x_F) \left[-1+\frac{4}{3}\gamma_E +\frac{2}{3} \ln\left( \frac{4 M^2(1-x_F)}{Q_0^2}  \right) + \frac{2}{3} \psi (5/2) \right] \Bigg. \\ \nonumber
\Bigg. +\frac{2}{3}   \int_{x_F}^1 dz \ \frac{\wp(x_F/z) [1+(1-z)^2] - \wp(x_F)}{1-z}
\Bigg\} .
\end{eqnarray}
Interestingly enough, after the $q_T$ integration a non-zero twist~3 contribution emerges for $\tilde{W}_{LT}$,
\begin{eqnarray}
\tilde{W}^{(3)}_{LT}=  \s0 \frac{ \sqrt{\pi} \left[2 -  \chi_{1}(3/2)- \chi_{2}(3/2)  \right]}{6}\frac{Q_0^3}{M^3} \wp(x_F).
\end{eqnarray}
This twist~3 emergence occurs in the perturbative part due to the presence of the $\sqrt{1-z}$ factor in (\ref{WtyldeLT}). It comes from the singular region, $z \to 1$, $q_T \to 0$. Since no twist~$3$ singularities are present in the color dipole cross-section, this contribution comes from a single pole in $s$ and carries no $\ln(M^2/Q_0)$. The twist~3 contributions to the other structure functions vanish, $\tilde{W}^{(3)}_{L} = \tilde{W}^{(3)}_{T}= \tilde{W}^{(3)}_{TT}=0$. It means, in particular, that the twist~3 contribution affects only the lepton angular distribution and not the total cross-section.

Expressions for twist 4 contributions in the integrated structure functions take the following form,
\begin{eqnarray}
 \tilde{W}^{(4)}_{L}&=&\frac{2}{15}\s0 \frac{Q_0^4}{M^4}\Bigg\{  \wp(x_F) \left[3-2\gamma_E - \ln\left( \frac{4 M^2(1-x_F)}{Q_0^2}  \right)   -\psi (7/2) \right] \Bigg.  \nonumber \\
& &
\Bigg. -  \int_{x_F}^1 dz \ \frac{\wp(x_F/z)z^2- \wp(x_F)}{1-z}
\Bigg\} , \\
 \tilde{W}^{(4)}_{T} & = &\s0 \frac{Q_0^4}{M^4}\Bigg\{  \frac{1}{30} \wp(x_F) \left[-46+24\gamma_E +12\ln\left( \frac{4 M^2(1-x_F)}{Q_0^2}  \right) +12\psi (7/2) \right] \Bigg.  \\
& &
 +\frac{1}{30} x_F \wp'(x_F) \left[17-12\gamma_E -6\ln\left( \frac{4 M^2(1-x_F)}{Q_0^2}\right) -7\psi (7/2) \right]+ \frac{1}{5} \frac{\wp(x_F)}{1-x_F}
\nonumber \\
& & \Bigg. -\frac{1}{5}  \int_{x_F}^1 dz \ \frac{\wp(x_F/z)[1+(1-z)^2] z^2- \wp(x_F)- (1-z)[2 \wp(x_F)-x_F\wp'(x_F)]}{1-z}
\Bigg\} , \nonumber \\
 \tilde{W}^{(4)}_{TT} & = & \s0 \frac{Q_0^4}{M^4}\Bigg\{ \frac{1}{60} \wp(x_F) \left[13-18\gamma_E + 30\ln4 -24\ln\left( \frac{4 M^2(1-x_F)}{Q_0^2} \right) +6\psi (7/2) \right] \Bigg.  \nonumber \\
& & \Bigg. -\frac{2}{5}  \int_{x_F}^1 dz \ \frac{\wp(x_F/z)z^2- \wp(x_F)}{1-z}
\Bigg\} , 
\end{eqnarray}
\begin{eqnarray}
 \tilde{W}^{(4)}_{LT} & = & \s0 \frac{Q_0^4}{16M^4}\Bigg\{ \frac{\pi  \wp(x_F)}{\sqrt{1-x_F}}-
\Bigg. \frac{\pi}{2}  \int_{x_F}^1 dz \ \frac{\wp(x_F/z)z^2(2-z)- \wp(x_F)}{(1-z)^{3/2}}
\Bigg\} .
\end{eqnarray}
The obtained results for the twist components of $\tilde W_T ^{(\tau)}$ and  $\tilde W_L ^{(\tau)}$ may be directly compared to the results of Ref.\ \cite{GBLS} after an integration of the lepton angles in the DY cross-section. This comparison was performed and the agreement was found. The contributions of two other structure functions $\tilde W_{TT} $ and $\tilde W_{LT}$ to the DY cross-section vanish after the integration over the lepton angles so they did not appear in the analysis of Ref.\ \cite{GBLS} and the twist decomposition of the $q_T$-integrated DY structure functions $\tilde W_{TT} $ and $\tilde W_{LT}$ is a novel result.

In accordance with Ref.\ \cite{GBLS} the obtained expression for twist-$\tau$ components of the DY integrated structure functions take a general form,
\begin{equation}
\sigma_0 \left( {Q_0 \over M} \right) ^{\tau} \,
\left[ \tilde A_1 ^{(\tau)} \ln \left({4M^2(1-x_F) \over Q_0^2 } \right) \,+\,
\tilde A^{(\tau)}_0+\tilde B^{(\tau)}
 \right],
\end{equation}
with the numerical coefficients $\tilde A^{(\tau)}_i$ given in the explicit analytic form and $\tilde B^{(\tau)}$ expressed as the integrals over $z$.
Both $\tilde A^{(\tau)}$ and $\tilde B^{(\tau)}$ depend on the fast quark density $\wp(x,M^2)$, in the projectile proton. The convergent integrals $\tilde B^{(\tau)}$ may be obtained by a numerical integration.
In general, the terms $\tilde B^{(\tau)}$ have no enhancement by $\ln (M^2/Q^2)$, so they are subleading, however they may be numerically important at a moderate $M^2$.

\subsection{The Lam-Tung relation}

One of the main goals of this paper is to prepare the ground for experimental measurements of the higher twist contributions in the forward DY scattering. Optimal observables for this purpose should have a suppressed leading twist contribution. One of such observables is given by the well known Lam-Tung relation
\cite{LamTung2}. Thus, within the parton model the following relation between the structure functions was proven \cite{LamTung2}:
\begin{equation}
T_1+\left ( \frac{q_P^2}{M^2}-1 \right)T_2-\frac{q_P q_p}{M^2}T_3+\left ( \frac{q_p^2}{M^2}+1 \right)T_4=0 .
\label{LT-Ti}
\end{equation}
Using (\ref{matrixl}) this can be rewritten as the equation for $W_i$ \cite{GJ}:
\begin{equation}
W_L-2 W_{TT}=0.
\label{LT-Wi}
\end{equation}
This relation holds in perturbative QCD at twist~2 up to the next-to-next-to-leading order correction, see e.g.\ \cite{GJ}.

Our results are consistent with the Lam-Tung relation. Indeed, from (\ref{WLtw2}) and  (\ref{WTTtw2}) one sees that $W^{(2)}_L-2 W^{(2)}_{TT}=0$. Moreover, as in Refs.\ \cite{FMSS,FSSM,GJ} we find a breakdown of the Lam-Tung relation at twist~4,
\begin{equation}
W^{(4)}_L-2 W^{(4)}_{TT}= \s0 \frac{Q_0^4}{M^4} \ \int_{x_F}^1 dz \  \wp(x_F/z) z^2 \frac{4M^{8} (1-z)^2}{\left[q_T^2+M^2 (1-z) \right]^4},
\label{LTrelTw4}
\end{equation}
and its integrated version,
\begin{eqnarray}
\int  \left(W^{(4)}_L-2 W^{(4)}_{TT} \right)  d^2 q_T= 2\pi \sigma_0 M^2 \left( \tilde{W}^{(4)}_L-2 \tilde{W}^{(4)}_{TT} \right) \\ \nonumber
=2\pi \sigma_0 \frac{Q_0^4}{M^2}\Bigg\{ \frac{1}{18} \wp(x_F) \left[-19+12\gamma_E+12\ln\left( \frac{ M^2(1-x_F)}{Q_0^2}  \right) \right] +\Bigg. \\ \nonumber
\Bigg. +\frac{2}{3}  \int_{x_F}^1 dz \ \frac{\wp(x_F/z)z^2- \wp(x_F)}{1-z}
\Bigg\} \, .
\end{eqnarray}
The breakdown term carries $\ln (M^2 / Q_0^2)$ so the Lam-Tung relation breaking at twist~4 enters at the leading order in QCD. Hence, the Lam-Tung combinations (\ref{LT-Ti}) and (\ref{LT-Wi}) of the forward DY structure functions exhibit an enhanced sensitivity to the higher twist effects and so they are promising
observables for the experimental finding of the higher twist effects at a small~$x$.

\section{Discussion and outlook}

The estimates of higher twist contributions to the Drell-Yan structure functions given in this paper are based on the eikonal GBW color dipole model cross-section. Although this model provides an efficient unified description of the small~$x$ DIS data down to the photoproduction limit, the diffractive DIS and the exclusive vector production, its twist content was not tested experimentally yet. In particular, the HERA data are consistent with the leading twist description, except of the kinematic edge of the very small~$x$ and moderate scales, and also with the Balitsky-Fadin-Kuraev-Lipatov (BFKL) and the Balitsky-Kovchegov (BK) cross-sections that have a different twist composition than the GBW model. Therefore, our results for the twist expansion of the forward DY structure functions are model-dependent predictions. Such an approach is justified by the lack of higher twist measurements at the small~$x$ domain, and the need to estimate the LHC potential to resolve the higher twist effects. On the other hand, clearly, for this latter goal one needs also to provide predictions dedicated for the LHC within other reasonable schemes, e.g.\ within the BFKL / BK approximation.

This paper is a necessary step towards such a broader analysis of the higher twists effects in the forward DY scattering at the LHC. It summarizes the key theoretical results needed for the higher twist extraction from the color dipole picture. In particular, the Mellin representation of the DY impact factors is a novel model independent result, following directly from perturbative QCD. This means in turn that certain features of the higher twist structure, following from properties of the impact factors, are generic, for example the breakdown of the Lam-Tung relation at twist~4. Also the presence or absence of $\ln (M^2 / Q_0^2)$ enhancement factor in the twist components of the structure functions is generic. It should be stressed that the obtained Mellin forms of the DY impact factors may be also used within the BFKL formalism for the small~$x$ resummation.

The next step in our program of the theoretical analysis of the LHC higher twist will be a data oriented study in which more models of the dipole cross-section will be considered and suitably refined to reproduce the existing forward DY data. Thus, we leave explicit numerical predictions for the higher twist corrections to forward DY scattering at the LHC to the second part of the analysis to be presented in the forthcoming paper \cite{MSS2}.

\paragraph{Acknowledgements}
Support of the Polish National Science Centre grants nos.\ DEC-2011/01/B/ST2/03643, DEC-2011/01/B/ST2/00492 and DEC-2014/13/B/ST2/02486 is gratefully acknowledged. TS acknowledges support in scholarship of Marian Smoluchowski Scientic Consortium Matter Energy Future from KNOW funding.

\appendix

\section{Derivation of relations between the helicity and the invariant structure functions}
\label{WvsT}

The invariant structure functions $T_i$ are defined as coefficients of hadron tensor decomposition \cite{LamTung1}:
\begin{eqnarray}
W^{\mu \nu}= -T_1\ \tilde{g}^{\mu \nu}+T_2\  \tilde{P}^{\mu}  \tilde{P}^{\nu} - T_3\ \frac{1}{2}\left( \tilde{P}^{\mu}  \tilde{p}^{\nu}+\tilde{p}^{\mu}  \tilde{P}^{\nu}\right) +T_4\ \tilde{p}^{\mu}  \tilde{p}^{\nu},
\label{invDef}
\end{eqnarray}
where $\tilde{g}^{\mu \nu}=g^{\mu \nu}-q^{\mu} q^{\nu}/q^2$, $P=P_1+P_2$, $p=P_1-P_2$ and $\tilde{P}^{\mu}=\tilde{g}^{\mu \nu} P_{\nu}/\sqrt{\sp}$, $\tilde{p}^{\mu}=\tilde{g}^{\mu \nu} p_{\nu}/\sqrt{\sp}$. The helicity structure functions are coefficients of hadron tensor decomposition using some coordinate system $(X^{\mu},Y^{\mu},Z^{\mu})$ in a lepton pair c.m.s.:
\begin{eqnarray}
W^{\mu \nu} &=& -\tilde{g}^{\mu \nu}(W_T+W_{TT}) -X^{\mu}  X^{\nu}  W_{TT}+ Z^{\mu}  Z^{\nu}(W_L-W_T-W_{TT})
\nonumber \\
& & - \left( \tilde{X}^{\mu}  Z^{\nu}+Z^{\mu}  X{\nu}\right) W_{LT}.
\label{helicityDef}
\end{eqnarray}
To find the relations between $W_i$ and $T_j$ one should relate  $(X,Y,Z)$ to $\tilde{P}, \ \tilde{p}$:
\begin{eqnarray}
Z^{\mu}=\alpha \tilde{P}^{\mu}+\beta \tilde{p}^{\mu},\;\;\;
X^{\mu}=\alpha' \tilde{P}^{\mu}+\beta' \tilde{p}^{\mu},
\label{ZXalphabeta}
\end{eqnarray}
where we assumed that the $Y$ axis is orthogonal to the reaction plane. Comparing (\ref{invDef}) and (\ref{helicityDef}) one finds:
\begin{eqnarray}
\label{matrixgenaral}
T_1&=&W_T+W_{TT},\\ \nonumber
T_2&=&-\alpha^2(W_{TT}-W_L+W_T)-2\alpha'(\alpha W_{LT}+\alpha' W_{TT}),    \\ \nonumber
T_3&=&2\alpha \beta(W_{TT}-W_L+W_T)+2\alpha'(\beta W_{LT}+2\beta' W_{TT})+2\alpha \beta'  W_{LT},  \\ \nonumber
T_4&=&-\beta^2(W_{TT}-W_L+W_T)-2\beta'(\beta W_{LT}+\beta' W_{TT}).
\end{eqnarray}

In our calculations the $Z$ axis is anti-parallel to the target momentum $\hat{Z}=-\vec {P}_1/|\vec{P}_1|$.
Since in the lepton pair c.m.s.\  $q^{\mu}=(M,0,0,0)$ one has $\tilde{P}^{\mu}=(0,\vec{P}/\sqrt{\sp})$, $\tilde{p}^{\mu}=(0,\vec{p}/\sqrt{\sp})$.
Comparing these two equations with (\ref{ZXalphabeta}) one arrives at the relation $\alpha=\beta$. Remembering that $X^2=Z^2=-1, \ X\cdot Z=0$
one obtains
\begin{eqnarray}
\alpha=\beta= -\frac{M}{q_P+q_p}, \;\;
\alpha'=\frac{M^2 q_P+q_T^2 q_p}{q_T(M^2+q_T^2)},\;\;
\beta'=-\frac{M^2 q_p+q_T^2 q_P}{q_T(M^2+q_T^2)},
\label{alphabeta}
\end{eqnarray}
where we denoted $q_P=q\cdot P/\sqrt{\sp}, \ q_p=q\cdot p/\sqrt{\sp}$. In the target rest frame $P_1=(m_p, 0, 0, 0)$ these scalar products become:
\begin{eqnarray}
q_P= \frac{x_F \sqrt{\sp}}{2}+\frac{M^2+q_T^2}{2x_F  \sqrt{\sp}},\;\;
q_p= \frac{x_F \sqrt{\sp}}{2}-\frac{M^2+q_T^2}{2x_F  \sqrt{\sp}} .
\label{qPqp}
\end{eqnarray}
Inserting (\ref{alphabeta}) with (\ref{qPqp}) into (\ref{matrixgenaral}) one obtains relations (\ref{matrixl}) between $T_j$ and $W_i$.

\section{The twist~2 components of the invariant structure functions}
\label{apTinv}

Applying the relations between the helicity and invariant DY structure functions one may find the 
twist decomposition of the invariant structure functions. We list the leading twist components of the DY invariant  structure functions:
\begin{eqnarray}
T_1^{(2)}&=&\sigma_0\frac{Q_0^2}{M^2} \int_{x_F}^1 dz \  \wp(x_F/z) \frac{M^4}{2 \left[q_T^2+M^2(1-z) \right]^4}
\left[4M^2q_T^2(1-z)^2+ \right.  \nonumber \\
& &
\left.+q_T^4(2-z(2-z))+M^4(1-z)^2 (2-(2-z)z) \right],
\\
T_2^{(2)}&=&\sigma_0\frac{Q_0^2}{M^2} \int_{x_F}^1 dz \  \wp(x_F/z)\frac{-M^6}{2s x_F^2 \left[q_T^2+M^2(1-z) \right]^4}
\left[2s^2x_F^4 (1-z)^2 \right.  \nonumber \\
& &
\left.+2sx_F^2 (q_T^2+M^2(1-z))z(1-z)+(q_T^2+M^2(1-z) )^2 z^2\right],
\\
T_3^{(2)}&=&\sigma_0\frac{Q_0^2}{M^2} \int_{x_F}^1 dz \  \wp(x_F/z) \frac{M^6}{2s x_F^2 \left[q_T^2+M^2(1-z) \right]^4}
\left[-2s^2x_F^4 (1-z)^2  \right.  \nonumber \\
& &
\left.+(q_T^2+M^2(1-z) )^2 z^2\right],
\\
T_4^{(2)}&=&\sigma_0\frac{Q_0^2}{M^2} \int_{x_F}^1 dz \  \wp(x_F/z) \frac{M^6}{2s x_F^2 \left[q_T^2+M^2(1-z) \right]^4}
\left[2s^2x_F^4 (1-z)^2 \right.  \nonumber \\
& &
\left.-2sx_F^2 (q_T^2+M^2(1-z))z(1-z)+(q_T^2+M^2(1-z) )^2 z^2\right].
\end{eqnarray}
Analogous expression for the higher twist components may be also obtained. They are however lengthy so we do not list them here.

\end{document}